\RequirePackage{ifpdf}
\ifpdf 
\documentclass[pdftex]{sigma}
\else
\documentclass{sigma}
\fi

\begin{document}
\allowdisplaybreaks

\renewcommand{\PaperNumber}{028}

\FirstPageHeading

\ShortArticleName{Representations of $U(2\infty)$ and the Value of
the Fine Structure Constant}

\ArticleName{Representations of $\boldsymbol{U(2\infty)}$ \\ and
the Value of the Fine Structure Constant}

\Author{William H. KLINK} \AuthorNameForHeading{W.H. Klink}

\Address{Department of Physics and Astronomy, University of Iowa,
Iowa City, Iowa, USA}
\Email{\href{mailto:william-klink@uiowa.edu}{william-klink@uiowa.edu}}

\ArticleDates{Received September 28, 2005, in final form December
17, 2005; Published online December 25, 2005}

\Abstract{A relativistic quantum mechanics is formulated in which
all of the interactions are in the four-momentum operator and
Lorentz transformations are kinematic.  Interactions are
introduced through vertices, which are bilinear in fermion and
antifermion creation and annihilation operators, and linear in
boson creation and annihilation operators. The fermion-antifermion
operators generate a unitary Lie algebra, whose representations
are fixed by a~first order Casimir operator (corresponding to
baryon number or charge).
 Eigenvectors and eigenvalues of the four-momentum operator
 are analyzed and exact solutions in the strong coupling limit
 are sketched.  A simple model shows how the fine
 structure constant might be determined for the QED vertex.}


\Classification{22D10; 81R10; 81T27}

\vspace{-2mm}

\section{Introduction}
With the exception of QCD and gravitational self couplings, all of
the fundamental particle interaction Hamiltonians have the form of
bilinears in fermion and antifermion creation and annihilation
operators times terms linear in boson creation and annihilation
operators.  For example QED is a theory bilinear in electron and
positron creation and annihilation operators and linear in photon
creation and annihilation operators.  The well-known
nucleon-antinucleon-meson interactions are of this form as are the
weak interactions.  These interactions differ of course in the way
the fermions are coupled to the bosons, including the way in which
internal symmetries are incorporated.  An exception is QCD, where
due to the $SU(3)_{\rm color}$ symmetry which generates gluon self
coupling terms, the gluon sector is no longer linear in creation
and annihilation operators. The other exception is the
gravitational interaction; since gravitons carry energy and
momentum, they also can couple to themselves.

If $a^{\dagger}$, $b^{\dagger}$ and $c^{\dagger}$ denote,
respectively, fermion, antifermion and boson creation operators,
the aformentioned interactions can all be written as
$(a^{\dagger}+b)(a+b^{\dagger})(c^{\dagger}+c)$, while the
``kinetic energy'' terms are of the form
$a^{\dagger}a-bb^{\dagger}+c^{\dagger}c$. Written in this way, the
fermion-antifermion bilinears $a^{\dagger}a$, $bb^{\dagger}$,
$a^{\dagger}b^{\dagger}$, and $ba$ close to form a Lie algebra
which, with indices attached to the creation and annihilation
 operators,  is related to the group $U(2N)$.
Similarly the boson operators $c^{\dagger}c$, $c^{\dagger}$ and
$c$ close to form a Lie algebra related to the semidirect sum of
unitary with  Heisenberg algebras.  Then the aforementioned
interactions can all be viewed as products of these two Lie
algebras.

The goal of this paper is to exploit this Lie algebra structure to
analyze eigensolutions of the four-momentum operator.  The
four-momentum operator $P^{\mu}$ will be written as the sum of
free and interacting four-momentum operators $P^{\mu}=P^{\mu}({\rm
fr})+P^{\mu}(I)$, where the free four-momentum operator has the
``relativistic kinetic energy'' form of creation and annihilation
operators, while the interacting four-momentum operator has the
Lie algebra product structure.

 To guarantee the relativistic
covariance of the theory, it is required that
\begin{gather}
[P^{\mu},P^{\nu}]=0,\label{eq1}\\
U_{\Lambda}P^{\mu}U_{\Lambda}^{-1}=\big(\Lambda^{-1}\big)^{\mu}_{\nu}P^{\nu},\label{eq2}
\end{gather}
where $U_{\Lambda}$ is the unitary operator representing the
Lorentz transformation $\Lambda$.  These ``point form''
equations~\cite{Klink 1998a}\footnote{For a discussion of the
various forms of dynamics, see for example~\cite{Keister 1991}.},
in which all of the interactions are in the four-momentum operator
and the Lorentz transformations are kinematic, lead to the
eigenvalue problem
\begin{gather}\label{eq3}
P^{\mu}|\Psi_p\rangle =p^{\mu}|\Psi_p\rangle,
\end{gather}
where $p^{\mu}$ is the four-momentum eigenvalue and
$|\Psi_p\rangle$ the eigenvector of the four-momentum operator,
which is an element in a generalized fermion-antifermion-boson
Fock space.   The physical vacuum and physical bound and
scattering states should then all arise as the appropriate
solutions of the eigenvalue equation \eqref{eq3}.  How this might
be done is sketched in the following sections.

\section{Vertex Interactions}
The free four-momentum operator is generated from irreducible
representations of the Poincar\'{e} group\footnote{The irreducible
representations of the Poincar\'{e} group are worked out in many
texts;  good examples of group theory texts are the texts by
A.O.~Barut and R.~R\c aczka~\cite{Barut 1977} and
W.~Tung~\cite{Tung 1985}.}.
 The positive mass,
positive spin representations act on a representation space that
can be written as the product of square integrable functions over
the forward hyperboloid times a~spin space,
$L^2(SO(1,3)/SO(3))\otimes V^j$, with the group action on states
given by
\begin{gather}\label{eq4}
U_{\Lambda}|v,\sigma\rangle=\sum|\Lambda v,\sigma^{'}\rangle
D^j_{\sigma',\sigma}(R_W),\qquad
 U_a|v,\sigma\rangle=e^{imv\cdot a}|v,\sigma\rangle,
\end{gather}
here $a$ is a four-vector space-time translation, $v$ is the
four-velocity on the forward hyperboloid satisfying $v\cdot v=1$
and $\sigma$ is the relativistic spin projection ranging between
$-j$ and $j$ and generating a basis in the $2j+1$ dimensional spin
space $V^j$. $m$ is a mass parameter that (along with the spin
$j$) labels an irreducible representation; the usual four-momentum
is defined by $p=mv$.  $D^j(\cdot)$ in equation~\eqref{eq4} is a
matrix element of the rotation group, with its argument a~Wigner
rotation defined by $R_W=B^{-1}(\Lambda v)\Lambda
B(v)$~\cite{Klink 1998}.

From this irreducible representation it is possible to define
many-particle creation and annihilation operators which satisfy
the following (anti)commutation relations:
\begin{gather}\label{eq6}
[a_{i},a^{\dagger}_{i'}]_{+}=\delta_{i,i'},\qquad
[b_{i},b^{\dagger}_{i'}]_{+}=\delta_{i,i'},\qquad 
[c_{k}, c^{\dagger}_{k'}]_-=\delta_{k,k'},
\end{gather}
the $a_i$, $b_i$, $c_k$ are respectively bare fermion, antifermion
and boson annihilation ope\-rators and satisfy appropriate
commutation or anticommutation relations with the associated
creation ope\-ra\-tors. They all transform in the same way as
one-particle states.  For example, $U_\Lambda
a^{\dagger}_{v,\sigma} U_\Lambda^{-1}=\sum a^{\dagger}_{\Lambda
v,\sigma'}D^j_{\sigma',\sigma}(R_W)$. The subscripts on these
operators denote whatever set of labels is inhe\-ri\-ted from the
one-particle representations.  Thus, if the fermions are labelled
by $v$ and $\sigma$, the Kronecker delta on the right hand sides
of equations~\eqref{eq6} stands for
$v_0\delta^3(v-v')\delta_{\sigma,\sigma'}$. It should be noted
that the three (anti)commutation relations are all dimensionless;
that is, by using the four-velocity rather than the four-momentum,
all quantities are dimensionless.

The free four-momentum operator can be written in terms of the
bilinears as
\begin{gather}\label{eq9}
P^{\mu}({\rm fr}):=m\sum \int dv\,
v^{\mu}\big(a^{\dagger}_ia_i+b^{\dagger}_i b_i+\kappa
c^{\dagger}_kc_k\big),
\end{gather}
where, as stated, $i$ stands for the collection of variables which
includes the four-velocity and other discrete variables such as
spin projections or internal symmetry labels.  $dv=\frac{d^3
v}{v_0}$ is the Lorentz invariant measure over the four-velocity,
which is common to all three types of particles. $k$ similarly
always contains the four-velocity, along with other discrete
variables that may differ from the fermionic variables. $\kappa$
is a dimensionless relative bare boson mass parameter which is
determined from solutions of the eigenvalue equation for physical
particles.  Thus, the only quantity with a dimension that appears
in equation~\eqref{eq9} is the bare fermion mass $m$; its value is
determined by relating a physical mass such as the nucleon mass to
the dimensionless eigenvalue of the corresponding stable particle.
Because of the transformation properties of the creation and
annihilation operators inherited from the one-particle states, the
free four-momentum operator as defined in equation~\eqref{eq9}
satisfies the point form equations~\eqref{eq1} and~\eqref{eq2}.

The interacting four-momentum operator is generated from vertices,
products of free field operators themselves made from creation and
annihilation operators.  If the vertex operator is denoted by
$V(x)$, where $x$ is a space-time point, then the interacting
four-momentum operator is obtained by integrating the vertex
operator over the forward hyperboloid:
\begin{gather}\label{eq10}
P^{\mu}(I):=g\int d^4 x\,\delta\big(x\cdot
x-\tau^2\big)\theta(x_0) x^{\mu} V(x)=g\int dx\, x^{\mu}V(x),
\end{gather}
where $g$ is a coupling constant and the measure $dx$ is defined
from equation \eqref{eq10}.  The vertex operator is required to be
a scalar density under Poincar\'{e} transformations and have
locality properties. That is,
\begin{gather}
U_a V(x) U_a^{-1}=V(x+a),\label{eq11}\\
U_\Lambda V(x)U_{\Lambda}^{-1}=V(\Lambda x),\label{eq12}
\end{gather}
$[V(x),V(y)]=0$ if $(x-y)^2$ is spacelike.  Here $U_a=e^{-iP({\rm
fr})\cdot a}$ is the free four-translation operator, with
$P^{\mu}({\rm fr})$ defined in equation~\eqref{eq9}.

Making use of the fact that if $x$ and $y$ are two time-like
four-vectors with the same length $\tau$ so that their difference
is space-like, it follows that $[P^{\mu}(I),P^{\nu}(I)]=0$.  Also,
from the Lorentz transformation properties of the vertex given in
equation~\eqref{eq12} it follows that the interacting
four-momentum operator transforms as a four-vector.  Thus
$P^{\mu}(I)$ also satisfies the point form equations \eqref{eq1}
and \eqref{eq2}.

However, what is important for a relativistic theory is that the
total four-momentum operator satisfy the point form equations.
Making the four-translations in equation~\eqref{eq11}
infinitesimal gives $[P^{\nu}({\rm fr}),P^{\mu}(I) ]=\int dx
\,x^{\mu}\frac{\partial}{\partial x_{\nu}}V(x)$ so that
\begin{gather*}
[P^{\nu}({\rm fr}),P^{\mu}(I)]-[P^{\mu}({\rm
fr}),P^{\nu}(I)]=g\int dx\left(x^{\mu} \frac{\partial}{\partial
x_{\nu}}-x^{\nu}\frac{\partial}{\partial
 x_{\mu}}\right)V(x)=0,\\
[P^{\mu}({\rm fr})+P^{\mu}(I), P^{\nu}({\rm fr})+P^{\nu}(I)]=[P^{\mu},P^{\nu}]=0.
\end{gather*}
Thus the total four-momentum operator also satisfies the point
form equations.

As stated in the Introduction, the vertex is assumed to be
bilinear in fermion-antifermion creation and annihilation
operators, and linear in boson creation and annihilation
operators.  That is,
$V(x)\sim(a^{\dagger}+b)(a+b^{\dagger})(c+c^{\dagger})$. For
example, for electromagnetic interactions the vertex is the
product of a local current operator times the photon field,
$V(x)=J^{\mu}(x)A_{\mu}(x)$, where the current operator is a
bilinear in fermion-antifermion creation and annihilation
operators (for example electron and positron creation and
annihilation operators) and the photon field is linear in photon
creation and annihilation operators.  Thus, the vertex operator
has the structure of a $U(2\infty)$ Lie algebra (which is analyzed
in the next section) times a Heisenberg-like algebra arising from
the boson part of the vertex.

\section[Representation structure of U(2N)]{Representation structure of $\boldsymbol{U(2N)}$}

The fermion-antifermion creation and annihilation operators appear
in the four-momentum ope\-rator only in the combinations
$a^{\dagger}_ia_j$, $b_ib^{\dagger}_j$,
$a^{\dagger}_ib^{\dagger}_j$ and $b_ia_j$. To analyze the
representation structure of such an algebra it is convenient to
temporarily assume the indices are discrete and range over $N$
values.  However, in the following sections the indices will not
be restricted to finite values.

To see how the four bilinears are related to a unitary algebra,
consider the following correspondence with  fermionic operators
$A_{\alpha}$, $A^{\dagger}_\beta$ ranging over $2N$ values:
$a^{\dagger}_i\rightarrow A^{\dagger}_i$, $a_i\to A_i$, $b_i\to
A^{\dagger}_{i+N}$, $b^{\dagger}_i\to A_{i+N}$, where bilinears in
the $A$'s and $A^{\dagger}$'s satisfy the unitary Lie algebra
(over $2N$~values)
 commutation relations
\begin{gather*}
[A^{\dagger}_{\alpha}A_{\beta},
A^{\dagger}_{\mu}A_{\nu}]=A^{\dagger}_{\alpha}A_{\nu}\delta_{\beta,\mu}
-A^{\dagger}_{\mu}A_{\beta}\delta_{\alpha,\nu}.
\end{gather*}

 For example
\begin{gather*}
[b_ia_j,
a^{\dagger}_kb^{\dagger}_l]=[A^{\dagger}_{i+N}A_j,A^{\dagger}_k
A_{l+N}]=
A^{\dagger}_{i+N}A_{l+N}\delta_{j,k}-A^{\dagger}_kA_j\delta_{i,l}
=b_ib^{\dagger}_l\delta_{j,k}-a^{\dagger}_ka_j\delta_{i,l},
\end{gather*}
as required.  It should be noted that if the antifermion creation
and annihilation operators are normal ordered, an extra term
(central extension) appears in the commutation relations. To avoid
such terms and keep the correspondence with the unitary algebra,
the antifermion creation and annihilation terms are not normal
ordered.

Define
\begin{gather}\label{eq17}
\mathcal{A}(X):=A^{\dagger}_{\alpha}X_{\alpha\beta}A_\beta, \qquad
0\le \alpha,\beta\le 2N,
\end{gather}
 then all fermionic terms in $P^\mu$ are of this
form.  For example, the free fermionic four-momentum operator can
be written, up to a constant, as
\begin{gather}
P^\mu_F({\rm fr})=m\sum\int dv\,v^\mu(a^{\dagger}_i
a_i-b_ib^{\dagger}_i) =m\sum\int\mathcal{A}(E^\mu),\qquad
E^\mu={\rm diag}\,(v^\mu,-v^\mu),\label{eq18}
\end{gather}
where in this expression continuous variables have been included.

Given the unitary algebra of $U(2N)$, representations can be given
in a variety of ways.  In the Gelfand--Zetlin scheme\footnote{A
good reference for the $U(N)$ groups, including the
Gelfand--Zetlin labeling scheme, is given in~\cite{Zelebenko
1973}.}, in which irreducible representations of the unitary
groups are labelled by nonnegative arrays of integers, the
relevant fermionic-antifermionic representations
 are all the antisymmetric representations, written
$(1,\dots,1,0,\dots,0)$, of length $2N$, with the identity
representations given by all zeroes or all ones.  The
antisymmetric Fock space is the direct sum of all these
irreducible representation spaces, and is of dimension $2^{2N}$.

When there are no internal symmetries, a more convenient way of
labelling the irreducible representations is with a ``baryon
number'' (or charge) operator.  Define $\mathcal{B}:=\sum\int
(a^{\dagger}_ia_i+b_ib^{\dagger}_i)=\sum\int\mathcal{A}(I)$, a
first order Casimir operator with eigenvalues from $-N$ to $+N$.
Each integer eigenvalue corresponds to a given irreducible
representation.  For example the $\mathcal{B}=0$ sector
corresponds in the Gelfand--Zetlin notation to arrays with equal
numbers of zeroes and ones.

For each baryon number sector there is a cyclic vector which is
annihilated by the lowering operators $ba$, and the raising
operators then generate all the possible states, up to a maximum
value, at which point the raising operator annihilates.  Such a
construction begins with the fermion-antifermion Fock vacuum,
$|0_F\rangle=|0\rangle|0\rangle$, the tensor product of fermion
and anti-fermion Fock vacuum.

Starting with baryon number zero, the fermion-antifermion cyclic
vector is defined by $a_i|0_F\rangle=b_i|0_F\rangle=0$ and hence
satisfies $b_ia_j|0_F\rangle=0$, $\mathcal{B}|0_F\rangle=0$.
States are generated from the vacuum by products of raising
operators
$|i_1j_1\cdots\rangle:=a^{\dagger}_{i_1}b^{\dagger}_{j_1}\cdots|0_F\rangle$.
The vacuum carries the identity representation for each $U(N)$
subalgebra, whereas the states
$|i,j\rangle:=a^{\dagger}_ib^{\dagger}_j|0_F\rangle$ carry the
tensor product of the fundamental representation of the $U(N)$
algebras.

Similarly for baryon number one, a cyclic vector is given by
$|0\rangle_1:=a^{\dagger}_i|0_F\rangle$, with
$b_ia_j|0\rangle_1=0$, $\mathcal{B}|0\rangle_1=|0\rangle_1$. Such
a cyclic vector carries the fundamental representation of the
$a^{\dagger}a$ algebra, and the identity representation of the
$b^{\dagger}b$ algebra.  A basis in the baryon number one irrep
space is again given by products of ``bare fermion-antifermion''
pairs
$|i_1j_1\cdots\rangle_1=a^{\dagger}_{i_1}b^{\dagger}_{j_1}\cdots|0\rangle_1$.
The dimension of this irrep space is smaller than the baryon zero
irrep space, because there are fewer ``fermion-antifermion'' pairs
that are available to span the space.

Continuing in this way there is finally the baryon $N$ irrep
space, a one dimensional space whose only state is the cyclic
vector itself, $|0\rangle_N:=a^{\dagger}_1\cdots
a^{\dagger}_N|0_F\rangle$, $\mathcal{B}|0\rangle_N=N|0\rangle_N$. The
other one dimensional irrep space is the baryon number $-N$ space,
defined by replacing the fermion creation operators with
anti-fermion creation operators.  These representations do not
exist when $N\rightarrow\infty$, which means they do not occur
when continuous variables are included.

\section{Eigenvalue structure of the four-momentum operator}

If the definition, equation~\eqref{eq17} is substituted into
equation~\eqref{eq10}, the interacting four-momentum operator can
be written as
\begin{gather}
P^\mu(I)=g\sum\int\big(\mathcal{A}(X^\mu_k)c_k+\mathcal{A}(X^\mu_k)^{\dagger}c^{\dagger}_k\big),\label{eq20}\\
[X^\mu_k,X^\nu_l]=0,\label{eq21}\\
[X^\mu_k,(X^\mu_l)^{\dagger}]=0.\label{eq22}
\end{gather}
Equations~\eqref{eq21} and \eqref{eq22} follow from the fact that
the components of the interacting four-momentum operator commute,
and that the four-momentum operator is hermitian.
Equation~\eqref{eq22} implies that the $X$'s are normal operators,
while equation~\eqref{eq21} implies that they can all be
simultaneously diagonalized.

The $(X^\mu_k)_{\alpha\beta}$ are obtained from a given vertex;
for example for the pseudoscalar coupling of a meson to nucleons
and antinucleons, two of the four terms in equation~\eqref{eq20}
are
\begin{gather*}
(X^\mu_k)_{ij}=F^\mu(v_1-v_2-v)\bar{u}(v_1,\sigma_1)\gamma_5u(v_2),\\
(X^\mu_k)_{i+N,j}=F^\mu(-v_1-v_2-v)\bar{v}(v_1,\sigma_1)\gamma_5u(v_2),\\
F^\mu(u):=\int d^4 y\,\delta(y\cdot y-1)\theta(y_0)y^\mu
e^{iu\cdot y}.
\end{gather*}

Since all the quantities in equation~\eqref{eq20} except the
coupling constant $g$ are dimensionless, $g$~has the dimensions of
mass.  If such a coupling constant is written as $g=m\alpha$,
where $m$ is the mass factor multiplying the free four-momentum
operator, equation~\eqref{eq9}, then $\alpha$ is a~dimensionless
coupling constant.  In the following the mass factor will be
dropped, so that the four-momentum operator is dimensionless.

Now the reason for writing the interacting four-momentum operator
in the form~\eqref{eq20}, is to solve the eigenvalue
problem~\eqref{eq3}.  If the one-particle eigenvalue problem can
be solved, the mass value of the eigenvalue will be dimensionless.
To connect it with a physical mass, the mass factor must be chosen
so that the physical mass equals the factor $m$ times the
dimensionless eigenvalue mass;  in this way the mass factor
multiplying both the free and interacting four-momentum operators
is fixed.

Combining the expressions for the terms in the four-momentum
operator given in equations~\eqref{eq9}, \eqref{eq20} the
eigenvalue problem~\eqref{eq3} to be solved can be written as
\begin{gather}\label{eq23}
\left(P^{\mu}_{F}({\rm fr})+\sum\int dv\big(\kappa
v^{\mu}c^{\dagger}_k c_k+\alpha \mathcal{A}(X^{\mu}_k)c_k+\alpha
\mathcal{A}(X^{\mu}_k)^{\dagger}c^{\dagger}_k\big)\right)|\Psi_{\lambda}\rangle
=\lambda^{\mu}|\Psi_{\lambda}\rangle,
\end{gather}
where the first term in equation~\eqref{eq23} is the free
fermion-antifermion four-momentum operator~\eqref{eq18}, the
second term is the boson free four-momentum operator, with the
factor~$\kappa$ the dimensionless ratio of physical meson mass to
physical fermion mass, and the third term is the interaction
coupling bilinear fermions to the mesons, with a strength given by
the dimensionless coupling constant~$\alpha$.  The point
eigenvalues will give the masses of the stable particles, while
the continuous part of the spectrum gives the scattering states.

Since the four-momentum operator transforms as a four-vector under
Lorentz transformations, Lorentz covariance can be used to write
$\lambda^{\mu}=(\lambda,0,0,0)$.  Then the eigenvector equation
can be rewritten as
\begin{gather*}
P^{0}_{F}({\rm fr})+\sum\int dv\big(\kappa v^{0}c^{\dagger}_k c_k
+\alpha\mathcal{A}(
X^{0}_k)c_k+\alpha\mathcal{A}(X^{0}_k)^{\dagger}c^{\dagger}_k\big)
|\Psi_{\lambda}\rangle =\lambda|\Psi_{\lambda}\rangle.
\end{gather*}
There is also a corresponding eigenvector equation for the space
components of the four-mo\-men\-tum operator, for which the
eigenvalue is zero.

Next consider an automorphism on the Heisenberg Lie algebra, with
transformation
\[
c_k=C_k-\frac{\alpha \mathcal{A}(X^0_k)^{\dagger}}{\kappa
v^0},\qquad c_k^{\dagger}=C_k^{\dagger}-\frac{\alpha
\mathcal{A}(X^0_k)}{\kappa v^0}.
\]
 This will be an automorphism if the $C$'s satisfy the correct
commutation relations,
\begin{gather*}
[C_k,C_{k'}^{\dagger}]=\left[c_k+\frac{\alpha\mathcal{A}(X^0_k)^{\dagger}}{\kappa
v^0},c_{k'}+\frac{\alpha \mathcal{A}(X^0_{k'})}{\kappa v^0}\right]
=\delta_{k,k'},\\ 
[C_k^{\dagger},C^{\dagger}_{k'}]=\left[c_k^{\dagger}+\frac{\alpha
\mathcal{A}(X^0_k)}{\kappa v^0}, c^{\dagger}_{k'}+\frac{\alpha
\mathcal{A}(X^0_{k'})}{\kappa v^0}\right]=0.
\end{gather*}
These commutation relations follow since the $\mathcal{A}(X)$'s
commute among themselves, equa\-tions~\eqref{eq21}.

Since the $C$'s also satisfy boson commutation relations, the
eigenvalue equation can be written in terms of them:
\begin{gather}\label{eq27}
\left(P^0_F({\rm fr})-\alpha^2\sum\int dv\,\frac{1}{\kappa
v^0}\mathcal{A}(X^0_k)\mathcal{A}(X^0_k)^{\dagger}+\sum\int
dv\,\kappa v^{0}C^{\dagger}_k C_k\right)|\Psi_{\lambda}\rangle
=\lambda|\Psi_\lambda\rangle.
\end{gather}
If $\alpha=0$ then the ground state solution is the usual Fock
space solution; but the above operators are all positive, so there
can never be a Fock space solution for $\alpha\neq 0$, that is,
when the physical vacuum is in the Fock space. The goal is to look
for solutions of equation~\eqref{eq27} that lead to a~suitable
choice for the physical vacuum.

\section{A simple model}
To motivate the kinds of considerations that go into getting a
physical vacuum, consider a simple model in which $N$ is finite
and the baryon number is $N$.  Then it is straightforward to get
exact solutions to the eigenvalue problem because the
fermion-antifermion sector is one-dimensional.

To get these solutions, it is simplest to realize the boson sector
in Bargmann variables, namely, $c_k^{\dagger}\rightarrow z_k$,
$c_k\rightarrow \frac{\partial}{\partial z_k}$ on a holomorphic
Hilbert space~\cite{Bargmann 1961}. Then the Hamiltonian and
eigenfunctions, eigenvalues, are
\begin{gather}
H=\sum
e_i\left(a^{\dagger}_ia_i+b^{\dagger}_ib_i+z_i\frac{\partial}{\partial
z_i}\right)+\alpha\sum\mathcal{A}(X_i^0)\frac{\partial}{\partial
z_i}+\mathcal{A}(X_i^0)^{\dagger}z_i\nonumber\\
\phantom{H}{}=\sum e_i+\sum\mathcal{A}(E)+\sum
e_i\left(z_i+\frac{\alpha\mathcal{A}(X_i^0)}{e_i}\right)\left(\frac{\partial}{\partial
z_i} +\frac{\alpha\mathcal{A}(X_i^0)^{\dagger}}{e_i}\right)\nonumber\\
\phantom{H=}{}
-\alpha^2\sum\mathcal{A}(X_i^0)\mathcal{A}(X_i^0)^{\dagger},\nonumber\\
|\Psi_n\rangle=\prod\left(e_kz_k+\frac{\alpha\mathcal{A}(X_k^0)}{e_k}\right)^{n_k}e^{-\alpha
\sum\frac{\mathcal{A}(X_i^0)z_i}{e_i}}a^{\dagger}_1\cdots a^{\dagger}_N|0\rangle, \label{eq30}\\
\lambda_n=\sum \left(e_i+n_ie_i-\frac{\alpha^2|Y_i^0|^2}{e_i}\right),\nonumber\\ 
\lambda_{\min}=\sum e_i-\alpha^2\sum\frac{|Y_i^0|^2}{e_i}\neq
0\label{eq32}.
\end{gather}
Here $e_i$ is the discretized energy, $(Y_i^0)=\sum
((X_i^0)_{jj}+(X_i^0)_{j+N,j+N})$, and $\kappa$ has been set to
one.  The minimum eigenvalue has a positive term coming from the
free ``relativistic energy'', balanced by a negative term, coming
from the interaction between fermions and bosons.

This simple model shows that the minimum eigenvalue expresed
by~\eqref{eq32} is not zero.  But the vacuum by definition has
zero energy.  As seen in~\eqref{eq32} only a definite value of the
coupling constant will satisfy such a requirement;  that is,
$H|\Psi_{\min}\rangle=0$ determines the value of the coupling
constant.

Now remove the $N$ finite restriction, replace the above
Hamiltonian with the electromagnetic Hamiltonian and replace the
baryon number operator with the charge operator in the charge zero
sector.  Since the free photon four-momentum operator has no
$\kappa$ factor (since photons are massless),  the vacuum solution
of $H$, namely $H|\Omega\rangle=0$, should fix the value of
$\alpha$, which gives the fine structure constant.  Note that
unlike the case of a Hamiltonian operator only, where it is always
possible to add an arbitrary constant to the Hamiltonian and hence
shift the energy by an arbitrary amount, in the case of the
four-momentum operator this is not possible without destroying the
Lorentz covariance of the operator.

A more interesting example is the above model in the $N=1$, baryon
number~0 sector.  Then the Hamiltonian and eigenfunctions for a
simple choice of $X$ can be written as
\begin{gather*}
H=e(a^{\dagger}a+b^{\dagger}b)+z\frac{\partial}{\partial z}
+\alpha(a^{\dagger}b^{\dagger}+ba)\left(\frac{\partial}{\partial
z}+z\right),\\
|\Psi_n\rangle
=f_1(z)|0\rangle+f_2(z)a^{\dagger}b^{\dagger}|0\rangle.
\end{gather*}

The eigenvalue problem generates a matrix differential equation
\begin{gather*}
z\frac{\partial f_1}{\partial z}+\alpha\frac{\partial
f_2}{\partial z}=
\lambda f_1-\alpha zf_2,\\
z\frac{\partial f_2}{\partial z}+\alpha\frac{\partial
f_1}{\partial z} =(\lambda -2e)f_2-\alpha zf_1
\end{gather*}
which can be rewritten as
\begin{gather*}
(z+\alpha)f_+'=(\lambda -e-\alpha z)f_++ef_-,\\
(z-\alpha)f_-'=(\lambda-e+\alpha z)f_-+ef_+
\end{gather*}
with $f_{\pm}:=f_1\pm f_2$, $f_{\pm}$ holomorphic. If $e$ is zero
the solution has a structure given by expression~\eqref{eq30}.
However, as far 
as the ``vacuum'' structure is concerned, $e$ not zero is the interesting 
case.

\section{Conclusion}

In the point form of relativistic quantum mechanics all interactions
are in the four-momentum operator, and Lorentz transformations are
kinematic (that is, Lorentz generators are free of interactions).
Since the four-momentum operator can be written as bilinear in
fermion, antifermion creation and annihilation operators, a Lie
algebra is generated, which is isomorphic to the Lie algebra of
the unitary group.  Further, the representations of interest are
the antisymmetric representations, and can be labelled by
integers, the eigenvalues of the baryon number operator.
Similarly the bosonic parts of the four-momentum operator are
elements of a Lie algebra, related to the semidirect sum of a
unitary algebra with the Heisenberg algebra.

Via the automorphism defined in Section~4, the four-momentum
operator can be rewritten in such a way as to exhibit properties
of its eigenvalue structure.  Though it is not discussed in this
paper, a strong coupling approximation of the four-momentum
operator, in which the fermion free relativistic  energy is
neglected, can be solved exactly and has a structure similar to
the solutions given in equation~\eqref{eq30}.

Several simple models show a balance of the minimum eigenvalue
between fermionic relativistic energy and fermion-boson coupling
terms;  this leads to the possibility that the fine structure
constant may be determined from requirements on the physical
vacuum.  However much remains to be done to connect the simple
models discussed here with  genuine infinite degree of freedom
systems.

\LastPageEnding

\end{document}